\documentclass[
 reprint,
 amsmath,amssymb,
 aps, prl,
]{revtex4-2}

\usepackage{appendix}
\usepackage{graphicx}
\usepackage{dcolumn}
\usepackage{bm}
\usepackage[hidelinks]{hyperref}
\usepackage[capitalise]{cleveref}
\usepackage{mhchem}
\usepackage{lipsum}
\usepackage[dvipsnames]{xcolor}
\usepackage{amsfonts}
\usepackage{yfonts}
\usepackage{nicematrix}
\NiceMatrixOptions{cell-space-limits = 2pt}

\usepackage[english]{babel}
\usepackage{amsthm}
\usepackage[scr=boondoxo,scrscaled=1.05]{mathalfa}
\theoremstyle{definition}

\DeclareFontFamily{U}{futm}{}
\DeclareFontShape{U}{futm}{m}{n}{
  <-> s * [.97534] fourier-bb 
  }{}
\DeclareMathAlphabet{\mathbbs}{U}{futm}{m}{n}

\begin{document}

\preprint{APS/123-QED}

\title{Nonequilibrium Macroscopic Response Relations for Counting Statistics}\author{Jiming Zheng}
\email{jiming@unc.edu}
\affiliation{Department of Chemistry, University of North Carolina-Chapel Hill, NC}
\author{Zhiyue Lu}
\email{zhiyuelu@unc.edu}
\affiliation{Department of Chemistry, University of North Carolina-Chapel Hill, NC}

\date{\today}

\begin{abstract}
Understanding how macroscopic nonequilibrium systems respond to changes in external or internal parameters remains a fundamental challenge in physics. In this work, we report a parameter transitional symmetry valid for macroscopic dynamics arbitrarily far from equilibrium. The symmetry leads to exact response relations and gives meaningful expansions in both linear and short-time regimes. This framework provides a universal description of macroscopic response phenomena arbitrarily far from equilibrium---including non-stationary processes and time-dependent attractors. The theory is validated and demonstrated numerically using the Willamowski–Rössler model, which exhibits rich dynamical behaviors including limit cycles and chaos.

\end{abstract}

\maketitle

{\it Introduction.---} Predicting the response of a nonequilibrium system to changes in its control or design parameters is a central challenge across physics, chemistry, and biology. At the microscopic scale, powerful tools exist: stochastic thermodynamics provides trajectory-level response relations and inequalities \cite{seifert2010fluctuation,baiesi2009nonequilibrium,baiesi2009fluctuations,maes2020response,zheng2025universal,zheng2025spatial,zheng2025nonlinear,kwon2025fluctuation}, while the linear-algebraic framework \cite{aslyamov2024general,aslyamov2025nonequilibrium,ptaszynski2024nonequilibrium,ptaszynski2025nonequilibrium} and matrix-tree descriptions \cite{owen2020universal,harunari2024mutual} of Markov jump processes yield various fluctuation-response relations. For macroscopic systems, however, such as chemical reaction networks (CRNs) \cite{vlad1994fluctuation,rao2016nonequilibrium,gaspard2020stochastic,lazarescu2019large} or many-body interacting systems \cite{meibohm2022finite,meibohm2024small,ptaszynski2025dissipation}, the state space of the master equation expands dramatically, rendering detailed microscopic information often inaccessible \cite{qian2021stochastic,falasco2025macroscopic}. Crucially, many natural and engineered systems operate on time-dependent attractors, including limit cycles and chaos—phenomena unique to macroscopic scales. Despite its fundamental importance, a general theory for the response of such macroscopic systems, from the bottom up, remains underdeveloped. Existing approaches are largely restricted to steady states \cite{vlad1994fluctuation,freitas2021linear,chun2023trade} or Gaussian noise approximations around fixed points \cite{aslyamov2025macroscopic}, leaving a universal macroscopic response relation elusive.

A systematic route from microscopic kinetics to emergent macroscopic dynamics is provided by the large deviation principle in the large-system-size limit \cite{qian2021stochastic,falasco2025macroscopic}. This framework yields an accurate description for general non-stationary and non-Gaussian fluctuating macroscopic processes. Crucially, it rigorously guarantees the correct minima of the dynamical rate function, which dictate the most probable macroscopic state. This foundational rigor contrasts Gaussian noise approximations, which can yield incorrect minima and lead to quantitatively flawed predictions for key physical quantities, such as entropy production \cite{gaveau1997master,hanggi1988bistability,vellela2009stochastic,gopal2022large}.

\begin{figure}[t]
    \centering
    \includegraphics[width=0.9\linewidth]{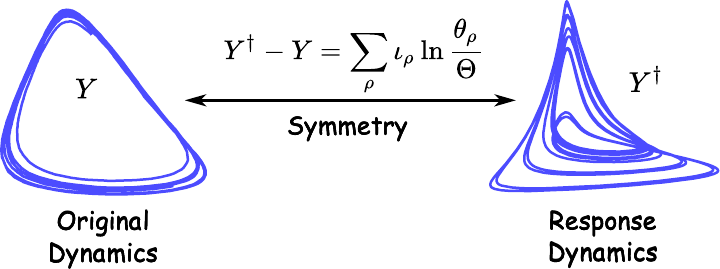}
    \caption{Parameter translational symmetry between macroscopic fluctuating trajectories generated with different dynamical parameters.}
    \label{fig: cover}
\end{figure}

Building upon this rigorous perspective, we identify a symmetry between the original and response macroscopic trajectories, as illustrated in \cref{fig: cover}. This symmetry serves as the cornerstone for an exact, finite-time macroscopic response theory for counting statistics under arbitrary parameter changes. Formulated at the level of macroscopic trajectories, our equality holds arbitrarily far from equilibrium, including in non-stationary processes and time-dependent attractors. Its expansions in the linear and short-time regimes further yield concise response relations. As a demonstration, we apply our theory to CRNs exhibiting limit cycles and chaos.

{\it From microscopic system to macroscopic system.---} Let us briefly review the macroscopic stochastic dynamics explained in detail in \cite{qian2021stochastic,falasco2025macroscopic}. Consider a system with $i = \{1, \cdots, N\}$ mesoscopic states. The vector of occupation numbers in each state $\boldsymbol{n} = (n_1, \cdots, n_N)^\top$ changes at random by a finite vector $\Delta_\rho = (\Delta_\rho^1, \cdots, \Delta_\rho^N)^\top$, where $\rho$ denotes different types of transitions. The probability distribution of $n_i$ at time $t$, denoted as $p(n_i, t)$, evolves according to the master equation:
\begin{equation}
    \frac{\partial p(n_i, t)}{\partial t} = \sum_\rho \left[ r_\rho(n_i-\Delta_\rho) p(n_i-\Delta_\rho, t) - r_\rho(n_i) p(n_i, t) \right],
    \label{eq: master equation}
\end{equation}
where $r_\rho(n_i)$ is the type-$\rho$ transition rate originating from $n_i$. Now we identify a large scaling parameter $V$ for the system, usually the volume, and consider the macroscopic limit: $\boldsymbol{n} \to \infty$, $V \to \infty$, and $\boldsymbol{c} \equiv \boldsymbol{n}/V < \infty$. We denote all the macroscopic quantities with uppercase letters. The transition rates scales as $R_\rho(\boldsymbol{c}) \equiv \lim_{V \to \infty} r_\rho(\boldsymbol{n}) / V$. The probability distribution of $c$ satisfies $P(\boldsymbol{c}, t) = V^N p(\boldsymbol{n}, t)$. The scaling relations lead to the macroscopic master equation:
\begin{equation}
    \frac{\partial_t P(\boldsymbol{c}, t)}{\partial t} = V\mathscr{H}(\boldsymbol{c}, -V^{-1}\partial_{\boldsymbol{c}}) P(\boldsymbol{c}, t),
    \label{eq: macroscopic master equation}
\end{equation}
where the Hamiltonian operator, i.e., the generator, is $\mathscr{H}(\boldsymbol{c}, -V^{-1}\partial_{\boldsymbol{c}}) = \sum_\rho \left[ e^{-V^{-1} \Delta_\rho \partial_{\boldsymbol{c}}} - 1 \right] R_\rho(\boldsymbol{c})$.
It requires that the probability distribution $P(\boldsymbol{c}, t)$ admits a large-deviation form $P(\boldsymbol{c}, t) \asymp e^{-V I(\boldsymbol{c}, t)}$, where $\asymp$ stands for asymptotically equal as $V\to \infty$ \cite{touchette2009large,dembo2009large} and $I(\boldsymbol{c}, t)$ is the large deviation rate function. This framework is powerful in describing various macroscopic nonequilibrium systems, including electronic systems \cite{freitas2022reliability,gopal2022large,freitas2021stochastic}, CRNs \cite{vlad1994fluctuation,rao2016nonequilibrium,gaspard2020stochastic,lazarescu2019large}, and many-body interacting systems \cite{meibohm2022finite,meibohm2024small,ptaszynski2025dissipation}.

For stochastic systems, observables are defined on trajectories $X_\tau$, which represent the time series of transitions occurring in the time interval $[0, \tau]$. A central class of trajectory observables are the macroscopic counting observables. These are defined as linear combinations of the scaled counts of specific transition events. For each transition type $\rho$, we define the macroscopic count as $\iota_{\rho}[X_{\tau}] = \lim_{V\to\infty} \frac{1}{V} m_{\rho}$, where $m_{\rho}$ is the total number of $\rho$-type transitions in the microscopic trajectory $X_{\tau}$. A general macroscopic counting observable is then given by $\iota[X_{\tau}] = \sum_{\rho} \omega_{\rho} \iota_{\rho}[X_{\tau}]$, where $\omega_{\rho}$ are fixed coefficients. Such observables are fundamental to many physics models and experiments, including diffusion \cite{levy1940certains,godreche2001statistics,majumdar2002local,lapolla2019manifestations,hartich2023violation}, active matter \cite{ramaswamy2010mechanics,ramaswamy2017active,marchetti2013hydrodynamics}, optics \cite{o2012time,margolin2005nonergodicity,ramesh2024arcsine,gopich2012theory}, chemical sensing \cite{bialek2005physical,endres2008accuracy,mora2010limits,govern2012fundamental,govern2014energy,harvey2023universal}, or biological transportation \cite{maffeo2012modeling,catterall2010ion,roux2004theoretical}. 

Now we introduce the counting field $\{ z_\rho \}$. General microscopic counting observables can be obtained from the moment generating function (MGF) $G(\{z_\rho\}, t) \equiv \langle e^{z_\rho m_\rho[X_\tau]} \rangle$, where $\langle \cdots \rangle$ is taken over all possible trajectories. Its macroscopic limit $\iota_\rho[X_\tau] \equiv \lim_{V\to\infty} \frac{m_\rho[X_\tau]}{V}$ follows the scaled cumulants generating function (SCGF) $K(\{z_\rho\}, t) \equiv \lim_{V\to\infty}\frac{1}{V}\ln G(\{z_\rho\}, t)$. The SCGF of any specific observable $\iota = \sum_\rho k_\rho \iota_\rho$, $K_{\iota}(z, t)$ , can be reduced by introducing $z_\rho = k_\rho z$. The MGF $G(\{z_\rho\}, t)$ has a path integral representation \cite{falasco2025macroscopic}:
\begin{equation}
    G(\{z_\rho\}, \tau) = \int\mathscr{D}c\int\mathscr{D}\pi e^{V\{ \mathscr{A}_{\{z_\rho\}}[\{c(t)\}, \{\pi(t)\}] - I(c(0), 0) \}},
\end{equation}
where $\mathscr{A}_{\{z_\rho\}}[\{c(t)\}, \{\pi(t)\}] = \int_0^\tau \mathrm{d}t \left[ -\pi  \frac{\mathrm{d} c}{\mathrm{d}t} + \mathscr{H}_{\{z_\rho\}}(c, \pi) \right]$ is the tilted path action, and $\mathscr{H}_{\{z_\rho\}}(c(t), \pi(t)) = \sum_\rho \left( e^{z_\rho + \pi\Delta_\rho} - 1 \right)R_\rho(c)$ is the tilted Hamiltonian. The observable distribution $P(\{Q_\rho\}, t)$ also takes the large-deviation form: $P(\{\iota_\rho\}, t) \asymp e^{-V Y(\{\iota_\rho\}, t)}$. The rate function $Y(\{\iota_\rho\}, t)$ can be obtained from the Legendre-Fenchel transform \cite{zia2009making}:
\begin{equation}
    Y(\{\iota_\rho\}, t) = \sup_{z} \left\{ K(\{z_\rho\}, t) - \sum_\rho z_\rho \iota_\rho \right\}.
    \label{eq: Legendre-Fenchel}
\end{equation}

{\it Macroscopic response relations.---} The transition rates $R_{\rho}$ typically depend on a set of intrinsic or extrinsic parameters, such as temperature, internal energy barriers, and external driving forces. When the parameter is changed at some time, the difference between the original and new dynamics is referred to as {\it the response of the system}. We denote the transition rates of the response dynamics by $R^\dagger_\rho$, and denote the ratio by $\theta_\rho \equiv R_\rho / R^\dagger_\rho$. To facilitate the derivation, we introduce a conjugate response dynamics by uniformly scaling all response rates: $R^{\ddagger}_{\rho} = \Theta R^{\dagger}_{\rho}$, where $\Theta \equiv \sum_{\rho}R_{\rho}/\sum_{\rho}R^{\dagger}_{\rho}$. This specific scaling ensures that the original and the conjugate dynamics share the same total escape rate: $\sum_{\rho}R_{\rho} = \sum_{\rho}R^{\ddagger}_{\rho}$. A key property of this construction is that the logarithmic ratio $\ln(\theta_{\rho}/\Theta)$ is odd under the conjugate transformation: $(\ln(\theta_{\rho}/\Theta))^{\ddagger} = -\ln(\theta_{\rho}/\Theta)$. With the conjugate response dynamics, we find the following symmetry of the tilted operator:
\begin{align}
    \mathscr{H}_{\{z_\rho\}}(c, \pi) &= \sum_\rho R_\rho \left( e^{\Delta_\rho \pi + z_\rho} - 1 \right) \nonumber \\
    &= \sum_\rho R^\ddagger_\rho \left( e^{\Delta_\rho \pi + z_\rho + \ln\frac{\theta_\rho}{\Theta}} -1 \right) \nonumber \\
    &= \mathscr{H}^\ddagger_{\left\{z_\rho-\ln\frac{\theta_\rho}{\Theta}\right\}}(c, \pi).
    \label{eq: Hamiltonian symmetry}
\end{align}
This operator symmetry \cref{eq: Hamiltonian symmetry} implies a corresponding symmetry for the tilted path action: $\mathscr{A}_{\{z_{\rho}\}}[\cdot] = \mathscr{A}^{\ddagger}_{\{z_{\rho}-\ln(\theta_{\rho}/\Theta)\}}[\cdot]$. Since the SCGF is determined by this action, it consequently obeys the symmetry: $K(\{z_{\rho}\},\tau) = K^{\ddagger}\left(\left\{z_{\rho}-\ln\frac{\theta_{\rho}}{\Theta}\right\},\tau\right)$. Combining it with the Legendre-Fenchel transform in \cref{eq: Legendre-Fenchel}, we obtain the finite-time detailed fluctuation relations between the original and conjugate response dynamics:
\begin{align}
    Y(\{\iota_\rho\}, \tau) &= \sup_{\{z_\rho\}} \left\{ K(\{z_\rho\}, \tau) - \sum_\rho z_\rho \iota_\rho \right\} \nonumber \\
    &= \sup_{\{z_\rho\}} \left\{ K^\ddagger \left( \left\{z_\rho-\ln\frac{\theta_\rho}{\Theta}\right\}, \tau \right) - \sum_\rho z_\rho \iota_\rho \right\} \nonumber \\
    &= Y^\ddagger(\{\iota_\rho\}, \tau) - \sum_\rho \iota_\rho \ln\frac{\theta_\rho}{\Theta}.
\end{align}

The conjugate response dynamics $\{ R^\ddagger_\rho \}$ scales all transition rates in response dynamics with the same constant $\Theta$. It is equivalent to measuring the response dynamics on another time scale \cite{di2018kinetic,hasegawa2023unifying}. Therefore, we have $Y^\ddagger(\{\iota_\rho\}, \tau) = Y^\dagger(\{\iota_\rho\}, \Theta\tau)$. With this equivalent representation, we obtain the fluctuation relations between the original and response dynamics:
\begin{equation}
    Y^\dagger(\{\iota_\rho\}, \Theta\tau) - Y(\{\iota_\rho\}, \tau) = \sum_\rho \iota_\rho \ln\frac{\theta_\rho}{\Theta}.
    \label{eq: macroscopic response relation}
\end{equation}
\cref{eq: macroscopic response relation} constitutes a macroscopic detailed fluctuation relation between original and response dynamics. It reveals that the log-ratio of probabilities for observing a given counting observable $\{\iota_{\rho}\}$ under the original and response dynamics is governed solely by a linear combination of the counts, weighted by the logarithmic change in the transition rates. This is our main result. It unravels a new symmetry between the original and response dynamics with arbitrarily large parameter change far from equilibrium or far from steady states. The r.h.s. of \cref{eq: macroscopic response relation} can be divided into two parts: $\sum_\rho \iota_\rho\ln\theta_\rho$ and $\sum_\rho \iota_\rho \ln\Theta$. The first part is solely contributed by the change in the dynamics, and the second part represents the effect of the change in systems' dynamical activity \cite{maes2020frenesy}. Increasing or decreasing the rate not only changes the trajectories' pattern but also modifies the dynamical activity. Our theory successfully captures the two effects and separates them into individual terms.

\cref{eq: macroscopic response relation} can also be written in an integral form:
\begin{equation}
    \left\langle \exp\left[ - V \sum_\rho \iota_\rho \ln\frac{\theta_\rho}{\Theta} \right] \right\rangle = 1.
    \label{eq: integral relation}
\end{equation}
Further applying Jensen's inequality leads to a {\it second-law-like} inequality for response parameters:
\begin{equation}
    \mathscr{R} \equiv \left\langle \sum_\rho \iota_\rho \ln\frac{\theta_\rho}{\Theta} \right\rangle \ge 0.
    \label{eq: second law like}
\end{equation}
Similar to the spirit of entropy production, one can use $\mathscr{R}$ to quantify how different the original and response dynamics are.

The derivation above assumes constant $\theta_{\rho}$ and $\Theta$. However, the theory generalizes directly to time-dependent parameters, which is crucial when the macroscopic rates $R_{\rho}(\boldsymbol{c})$ and $R^{\dagger}_{\rho}(\boldsymbol{c})$ depend explicitly on the instantaneous concentrations $c(t)$. In this general case, the macroscopic response relation \cref{eq: macroscopic response relation} extends to:
\begin{equation}
    Y^{\dagger}\left(\{\iota_{\rho}\},\int_{0}^{\tau}\Theta dt\right) - Y(\{\iota_{\rho}\},\tau) = \sum_{\rho} \int_{0}^{\tau} \dot{\iota_\rho}(t) \ln\frac{\theta_{\rho}}{\Theta} dt,
\end{equation}
where $\dot{\iota_\rho}$ is the time-derivative of $\iota_\rho$. For time-dependent cases, \cref{eq: integral relation,eq: second law like} are modified accordingly.

{\it Linear response regime.---} The response relation in \cref{eq: macroscopic response relation} is inherently nonlinear. To access the linear response regime, we consider the case of a small parameter perturbation. To derive the linear response, we consider an infinitesimal perturbation $\mathrm{d} R_{\rho}$ to each rate and expand the left-hand side of \cref{eq: macroscopic response relation} to first order. Performing this expansion around the unperturbed system yields (see Appendix A):
\begin{equation}
    \sum_{\rho}\frac{\partial Y(\{\iota_{\rho}\},\tau)}{\partial\ln R_{\rho}}\cdot\Delta R_{\rho} = -\sum_{\rho} \iota_{\rho} R_{\rho},
    \label{eq: linear inner product}
\end{equation}
where $\Delta R_\rho \equiv \frac{\mathrm{d} R_\rho}{R_\rho} - \frac{\sum_{\rho'} \mathrm{d} R_{\rho'}}{\sum_{\rho'} R_{\rho'}}$ and $\mathrm{d} R_\rho$ is the infinitesimally small change in the transition rate $R_\rho$. This specific form of $\Delta R_{\rho}$ ensures that the perturbation does not alter the total escape rate, $\sum_{\rho} R_{\rho} \Delta R_{\rho} = 0$, by construction. We also justify that $\{R_\rho\}$ is the only vector perpendicular to $\{\Delta R_\rho\}$ \footnote{Because of the condition $\sum_\rho R_\rho \Delta R_\rho = 0$, $\Delta R_\rho$ has one less degree of freedom than $\mathrm{d} R_\rho$. The space perpendicular to $\Delta R_\rho$ is one-dimensional. Therefore, the vector $\{ R_\rho \}$ we already found is the only one perpendicular to it.}. Consequently, the solution to the linear response equation must take the form:
\begin{equation}
    \frac{\partial Y(\{\iota_{\rho}\},\tau)}{\partial\ln R_{\rho}} = -\iota_{\rho} + C R_{\rho},
    \label{eq: rate function linear response}
\end{equation}
where $C$ is a quantity that depends on the trajectory but is independent of $R_\rho$. It states that the linear response $\partial_{\ln R_{\rho}}Y(\{\iota_\rho\}, \tau)$ is linearly dependent on $R_\rho$. For general parameters, the response is given by the chain rule expansion. For steady-state systems, we derive the explicit expression in Appendix B:
\begin{equation}
    \frac{\partial Y(\{\iota_\rho\}, \tau)}{\partial \ln R_\rho} = - \iota_\rho + R_\rho \frac{\sum_{\rho'} \iota_{\rho'}}{\sum_{\rho'} R_{\rho'}}
    \label{eq: rate function linear response}
\end{equation}
with $C = \sum_{\rho'} \iota_{\rho'} / \sum_{\rho'} R_{\rho'}$. Due to probability conservation, the trajectory average of $\partial_{\ln R_{\rho}}Y(\{\iota_\rho\}, \tau)$ must be zero. Therefore, we obtain
\begin{equation}
    \frac{\langle \iota_\rho \rangle}{R_\rho} = \frac{\sum_{\rho'} \langle \iota_{\rho'} \rangle}{\sum_{\rho'} R_{\rho'}}.
    \label{eq: steady-state relation}
\end{equation}
This implies that the mean current divided by rate, $\langle \iota_\rho \rangle / R_\rho$, has the same value for all transition channel $\rho$ in macroscopic steady state. Later, we show that \cref{eq: steady-state relation} offers a convenient way to verify our theory numerically.

{\it Short-time limit.---} The effect of $\Theta$ is coupled with $Y^\dagger$ as $Y^\dagger(\{\iota_\rho\}, \Theta\tau)$. To separate them, we consider the short-time limit of our theory. The expansion around $\tau = 0$ gives the short-time response relation
\begin{equation}
    \Theta \frac{\partial Y^\dagger(\{\iota_\rho\}, 0)}{\partial \tau} - \frac{\partial Y(\{\iota_\rho\}, 0)}{\partial \tau} = \sum_\rho \dot{\iota_\rho} \ln \frac{\theta_\rho}{\Theta}.
\end{equation}
We further use the inequality $\Theta \le \sum_\rho \theta_\rho \equiv \hat{\theta}$. It yields the upper bound on response in the short-time regime:
\begin{equation}
    \hat{\theta} \frac{\partial Y^\dagger(\{\iota_\rho\}, 0)}{\partial\tau} - \frac{\partial Y(\{\iota_\rho\}, 0)}{\partial\tau} \le \sum_\rho \dot{\iota_\rho} \ln\frac{\theta_\rho}{\hat{\theta}}.
\end{equation}
In the short-time inequality, only intrinsic rate parameters appear. For example, for chemical reactions, the short-time ratio $\theta$ is the ratio between reaction constants $k_\rho / k^\dagger_\rho$ and solely depends on the parameter change. If the parameter is changed by applying the driving force $F$ to the reaction $\rho$ with $k^\dagger_\rho = k_\rho e^{F}$ and $k^\dagger_{-\rho} = k_{-\rho} e^{-F}$, then the short-time ratio is $\theta_{\pm\rho} = e^{\mp F}$. For unchanged rates, $\theta_\rho$ generally contributes $1$.

Now, consider systems initially prepared at steady states. The time derivative $\partial_\tau Y(\{\iota_\rho\}, 0)$ is zero. In this case, the short-time response of the steady-state system is bounded from above as
\begin{equation}
    \frac{\partial Y^\dagger(\{\iota_{\rho}\}, 0)}{\partial \tau} = \frac{1}{\Theta} \sum_\rho \dot{\iota_\rho}\ln\frac{\theta_\rho}{\Theta} \le \frac{1}{\hat{\theta}} \sum_\rho \dot{\iota_\rho}\ln\frac{\theta_\rho}{\hat{\theta}}.
\end{equation}
Crucially, the right-hand side of this inequality depends solely on the intrinsic parameter ratios $\theta_\rho$ and the initial steady-state currents $\dot{\iota_{\rho}}$, and is independent of the specific instantaneous concentrations. Notice that this inequality applies to arbitrarily large parameter changes.

{\it Demonstration.---} We take CRNs as examples to illustrate and verify our theory. Here, we apply our theory to the minimal Willamowski-R\"ossler model of chemical reactions \cite{rossler1976chaotic,bodale2015chaos}. It is well-known for exhibiting rich dynamics—including fixed points, limit cycles, and chaos—under different parameter sets. The model contains the following elementary reactions:
\begin{subequations}
\begin{gather}
    \ce{A_1 + X <=>[\text{$k_1$}][\text{$k_{-1}$}] 2X};  \quad \ce{X + Y ->[\text{$k_2$}] 2Y}; \nonumber \\
    \ce{A_4 + Y ->[\text{$k_3$}] A_2}; \quad \ce{X + Z ->[\text{$k_4$}] A_3}; \nonumber \\
    \ce{A_5 + Z <=>[\text{$k_5$}][\text{$k_{-5}$}] 2Z}; \nonumber
\end{gather}
\end{subequations}
where the concentrations of species $\{\ce{A}_i\}$ are maintained constant with particle reservoirs. In this case, the mesoscopic states are $\{\ce{X}, \ce{Y}, \ce{Z}\}$, the occupation numbers and concentrations are $(n_{\ce{X}}, n_{\ce{Y}}, n_{\ce{Z}})$ and $(c_{\ce{X}}, c_{\ce{Y}}, c_{\ce{Z}})$, respectively. The transition $\rho$ refers to the seven reactions (five forward and two backward), and the numbers of particles that change in each reaction are denoted by the transition vector $\Delta_\rho$.

\begin{figure}[htbp]
    \centering
    \includegraphics[width=0.95\linewidth]{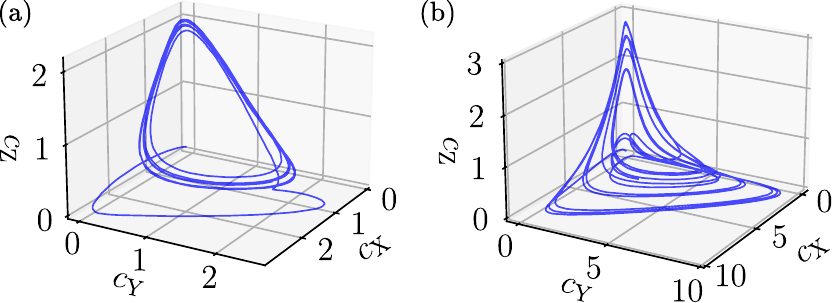}
    \caption{Stochastic trajectories of the Willamowski-R\"ossler model. (a) Limit cycle attractor. (b) Chaotic (strange) attractor.}
    \label{fig: attractor}
\end{figure}

In the following simulations, we set the volume $V$ to be $10^5$. The concentrations $\boldsymbol{c} = \boldsymbol{n}/V$ are $0.1$ for all $\{\ce{A}_i\}$. Initial concentrations are $c_{\ce{X}} = 0.21$, $c_{\ce{Y}} = 0.01$, and $c_{\ce{Z}} = 0.12$. The reaction constants are $k_1 = 30$, $k_2 = 1$, $k_3 = 10$, $k_4 = 1$, $k_5 = 16.5$, and $k_{-5} = 0.5$. The system exhibits a limit cycle when $k_{-1} = 1$ and chaotic behavior when $k_{-1} = 0.25$, as shown in \cref{fig: attractor}. All simulations are based on the Gillespie algorithm on \cref{eq: master equation}.

\begin{figure*}[htbp]
    \centering
    \includegraphics[width=0.8\linewidth]{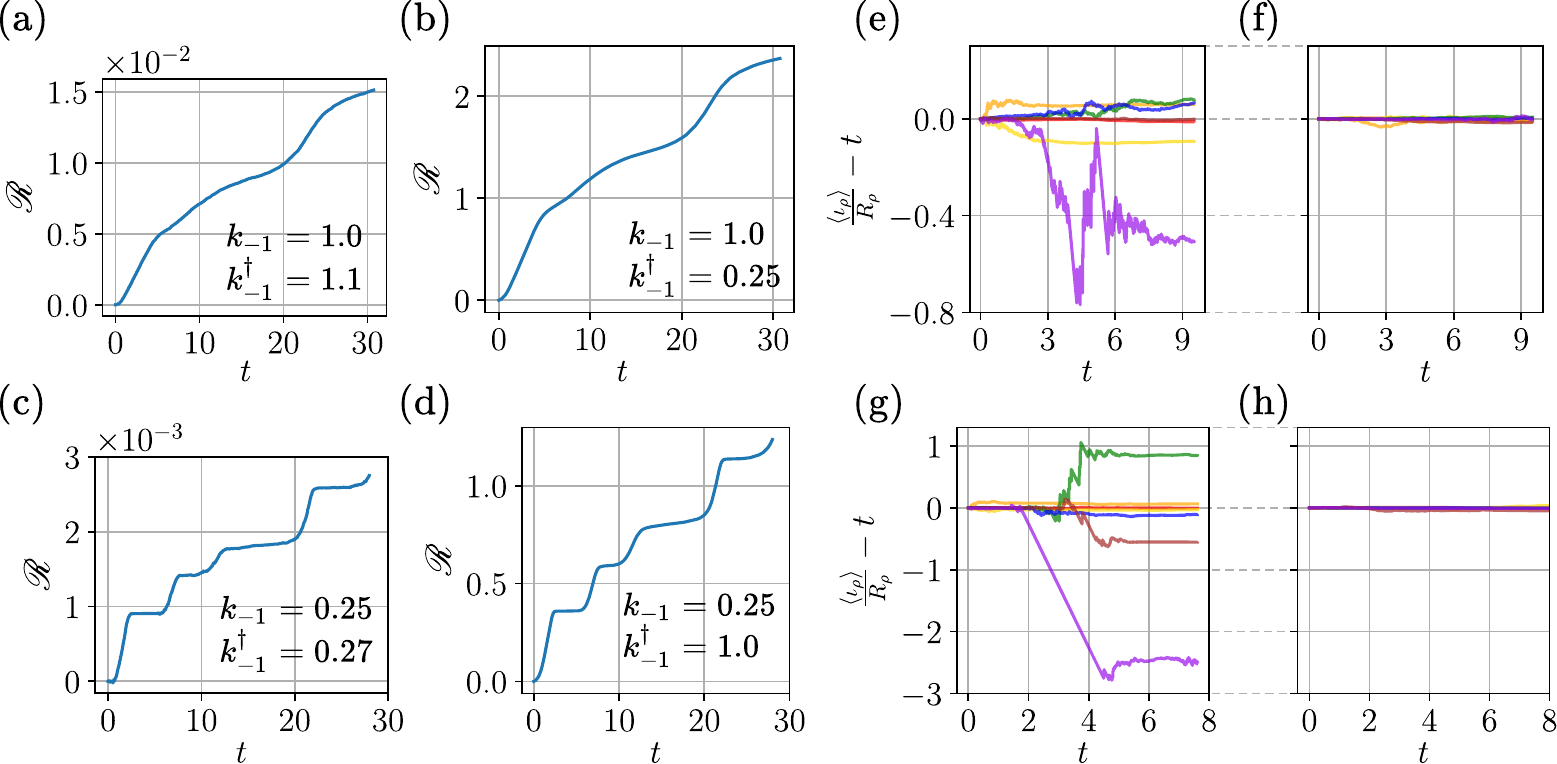}
    \caption{Numerical validation of the macroscopic response relations. (a-d) The positivity of the response measure $\mathscr{R}(t)$ (\cref{eq: second law like}) for limit cycle (a,b) and chaotic (c,d) attractors under different perturbations. (e-h) The time evolution of the normalized mean currents $\langle\iota_{\rho}\rangle/R_{\rho}$ (\cref{eq: steady-state relation}), showing convergence to a common value $t$ in the steady state for both limit cycles (e,f) and chaos (g,h). (f) and (h) are obtained after pre-equilibration time $t_\text{pre-eq} = 10$. Color legend: red for $\rho = 1$, orange for $\rho = 2$, yellow for $\rho = 3$, green for $\rho = 4$, blue for $\rho = 5$, brown for $\rho = -1$, purple for $\rho = -5$.}
    \label{fig: numerical}
\end{figure*}

Direct evaluation and visualization of \cref{eq: macroscopic response relation} is challenged by the exponential suppression of rare events in the macroscopic limit and the high dimensionality of the rate function. We therefore test \cref{eq: second law like,eq: steady-state relation} on the stochastic trajectories. We compute the response quantity $\mathscr{R}$ for both limit cycle and chaotic attractors under different parameter perturbations. As shown in \cref{fig: numerical}(a-d), $\mathscr{R}(t)$ remains non-negative at all times. The temporal profile of $\mathscr{R}(t)$ is dictated by the underlying attractor dynamics, while its magnitude scales with the extent of the parameter change. We then examine the normalized mean currents $\langle\iota_{\rho}\rangle/R_{\rho}$. When the system evolves from a non-stationary initial state (\cref{fig: numerical}e, g), these quantities relax toward a common value. Crucially, once the system reaches its steady state (\cref{fig: numerical}f, h), all $\langle\iota_{\rho}\rangle/R_{\rho}$ collapse onto a single curve, confirming \cref{eq: steady-state relation}. Furthermore, this universal value is numerically indistinguishable from the total trajectory time $t$, leading us to the general conclusion that $\langle\iota_{\rho}\rangle / R_{\rho} = t$ holds for any transition channel $\rho$ in a macroscopic steady state, regardless of whether it is a fixed point, limit cycle, or chaotic attractor.

{\it Discussion.---} We discuss the connections between our result and trajectory-based response theories for Markov dynamics. The result in this letter elegantly connects to the microscopic trajectory-level response relation $\mathcal{P}[X_\tau]$, which is written as $\frac{\partial \ln \mathcal{P}[X_\tau]}{\partial \ln r_{ij}} = m_{ij} - r_{ij}\tau_j$ \cite{seifert2010fluctuation,zheng2025spatial,zheng2025nonlinear}, where $\mathcal{P}[X_\tau]$ is the trajectory probability, $r_{ij}$ is the microscopic transition rate from state $j$ to state $i$, $m_{ij}$ is the number of transitions, and $\tau_j$ is the total occupation time on the state $j$. Our macroscopic linear response relation \cref{eq: rate function linear response} emerges as the limit of this microscopic formula. The passage to the macroscopic limit, however, is highly non-trivial. As the system size grows, the discrete occupation times $\tau_{j}$ for individual microstates become ill-defined in the continuous concentration space, while the number of accessible states diverges. Our theory resolves this scaling issue by proving the exists of a well-defined limit and providing its explicit form in terms of macroscopic steady-state currents. From the microscopic point of view, we can further claim $\langle \iota_\rho \rangle/R_\rho = t$. The macroscopic rate $R_\rho$ is composed of all microscopic rates $r_\rho$ sharing the same transition vector $\Delta_\rho$. Therefore, the occupation time $\tau_j$ involves all states, excluding those on the boundary (where the concentration of at least one species is zero). As a result, the macroscopic limit of $\langle \tau_j \rangle$ for any $\rho$ is the total time length.

There is also a response fluctuation relation for microscopic master equations \cite{maes2020response}. It states that the microscopic trajectory-level response is related to the system's change in entropy production and dynamical activity. In our framework, the contribution of dynamical activity is captured by $\sum_\rho \iota_\rho \ln\Theta$. An open and significant question is whether the remaining term $\sum_\rho \iota_\rho \ln \theta_\rho$ is generally related to entropy production. It would be interesting to further explore the detailed connections between the two in the future.

Ultimately, It is important to distinguish our macroscopic response relation from conventional fluctuation theorems. Typical fluctuation theorems on currents \cite{crooks1999entropy,andrieux2004fluctuation,andrieux2006fluctuation,andrieux2007fluctuation} typically establish a symmetry between the probability of a forward trajectory and its time-reversed counterpart. Conventional fluctuation theorems fundamentally relies on micro-reversibility, requiring the underlying dynamics to be physically reversible (i.e., if there is a forward rate, there is a backward rate). In contrast, our theory compares the probabilities of the same forward trajectory under two different sets of parameters. This reveals a new symmetry with respect to translations in the parameter space, rather than time reversal. Consequently, our theory does not require the transition rates to be reversible and is therefore applicable to a broad class of systems arbitrarily far from equilibrium.

{\it Conclusion.---} In this Letter, we establish a fundamental symmetry between original and response dynamics for macroscopic stochastic systems, which we formulate as a response relation for counting observables. This relation provides a universal description of nonequilibrium response, valid for finite times and arbitrarily far from equilibrium---including non-stationary processes and time-dependent attractors. The linear expansion of our theory yields a macroscopic linear response relation, while its short-time limit produces inequalities independent of instantaneous states. We numerically verified our theory using the Willamowski-Rössler model, demonstrating its validity for systems exhibiting limit cycles and chaotic attractors. Our framework thus offers a powerful and general tool for probing responses in a wide range of macroscopic nonequilibrium phenomena.

{\it Acknowledgments.---} This work is supported by the National Science Foundation under Grant No. DMR-2145256 and Alfred P. Sloan Foundation Award under grant number G-2025-25194.

\appendix
\section{End Matter}
\subsection{Appendix A: Derivation of \cref{eq: linear inner product}} \label{SIsec: linear expansion}
\setcounter{equation}{0}
\renewcommand{\theequation}{A\arabic{equation}}
In the derivation, we use $Y(\tau)$ as an abbreviation of $Y(\{\iota_\rho\}, \tau)$. For $R^\dagger_\rho = R_\rho + \mathrm{d}R_\rho$, the quantity $\Theta \equiv \sum_\rho R_\rho / \sum_\rho R^\dagger_\rho$ can be expanded as
\begin{subequations}
\begin{align}
    \Theta &= \frac{\sum_\rho R_\rho}{\sum_\rho R_\rho + \sum_\rho \mathrm{d}R_\rho} \\
    &= \frac{1}{1 + \sum_\rho \mathrm{d}R_\rho / \sum_\rho R_\rho} \\
    &= 1 - \frac{\sum_\rho \mathrm{d}R_\rho}{\sum_\rho R_\rho} + o(\mathrm{d}R_\rho).
\end{align}
\end{subequations}
The left-hand side of \cref{eq: macroscopic response relation} can be expanded as
\begin{subequations}
\begin{align}
    &Y^\dagger(\Theta\tau) - Y(\tau) \nonumber \\
    ={}& Y^\dagger(\tau) - \tau \frac{\partial Y^\dagger(\tau)}{\partial\tau}\frac{\sum_\rho \mathrm{d}R_\rho}{\sum_\rho R_\rho} - Y(\tau) + o(\mathrm{d}R_\rho) \\
    ={}& \sum_\rho \frac{\partial Y(\tau)}{\partial R_\rho} \mathrm{d}R_\rho - \tau \frac{\partial Y^\dagger(\tau)}{\partial\tau}\frac{\sum_\rho \mathrm{d}R_\rho}{\sum_\rho R_\rho} + o(\mathrm{d}R_\rho) \\
    ={}& \sum_\rho \frac{\partial Y(\tau)}{\partial \ln R_\rho} \frac{\mathrm{d}R_\rho}{R_\rho} - \tau \frac{\partial Y(\tau)}{\partial\tau}\frac{\sum_\rho \mathrm{d}R_\rho}{\sum_\rho R_\rho} + o(\mathrm{d}R_\rho).
    \label{SIeq: general expansion}
\end{align}
\end{subequations}
Expanding the time-scale relation $Y^\ddagger(\Theta\tau) = Y^\dagger(\tau)$ yields
\begin{equation}
    \sum_\rho \frac{\partial Y(\tau)}{\partial \ln R_\rho} = \tau \frac{\partial Y(\tau)}{\partial \tau},
\end{equation}
which is already known in \cite{di2018kinetic}. Therefore, the left-hand side of \cref{eq: macroscopic response relation} becomes
\begin{equation}
    Y^\dagger(\Theta\tau) - Y(\tau) = \sum_\rho \frac{\partial Y(\tau)}{\partial \ln R_\rho} \cdot \Delta R_\rho + o(\mathrm{d}R_\rho),
\end{equation}
where $\Delta R_\rho = \frac{\mathrm{d}R_\rho}{R_\rho} - \frac{\sum_\rho \mathrm{d}R_\rho}{\sum_\rho R_\rho}$. It is straightforward to check that $\sum_\rho R_\rho \Delta R_\rho = 0$.

The log-factor on the right-hand side of \cref{eq: macroscopic response relation} can be expanded as
\begin{subequations}
\begin{align}
    \ln\frac{\theta_\rho}{\Theta} &= \ln \theta_\rho - \ln \Theta \\
    &= \ln\frac{1}{1 + \mathrm{d}R_\rho/R_\rho} - \ln\frac{1}{1 + \sum_\rho \mathrm{d}R_\rho / \sum_\rho R_\rho} \\
    &= \frac{\sum_\rho \mathrm{d}R_\rho}{\sum_\rho R_\rho} - \frac{\mathrm{d}R_\rho}{R_\rho} + o(\mathrm{d}R_\rho).
\end{align}
\end{subequations}
Therefore, the right-hand side of \cref{eq: macroscopic response relation} becomes
\begin{equation}
    \sum_\rho \iota_\rho \ln\frac{\theta_\rho}{\Theta} = - \sum_\rho \iota_\rho \Delta R_\rho + o(\mathrm{d}R_\rho).
\end{equation}
As a result, we obtain \cref{eq: linear inner product}.

\subsection{Appendix B: Derivation of \cref{eq: rate function linear response}}
\setcounter{equation}{0}
\renewcommand{\theequation}{B\arabic{equation}}

For stationary states, the shape of the rate function do not change with time. So $\partial Y(\tau)/\partial\tau = 0$. In this case, \cref{SIeq: general expansion} becomes
\begin{equation}
    Y^\dagger(\Theta\tau) - Y(\tau) = \sum_\rho \frac{\partial Y(\tau)}{\partial \ln R_\rho} \frac{\mathrm{d}R_\rho}{R_\rho} + o(\mathrm{d}R_\rho).
\end{equation}
It leads to the steady-state linear response relation:
\begin{equation}
    \sum_\rho \frac{\partial Y(\tau)}{\partial \ln R_\rho} \frac{\mathrm{d}R_\rho}{R_\rho} = \sum_\rho \left[ \iota_\rho \left( \frac{\sum_{\rho'} \mathrm{d}R_\rho}{\sum_{\rho'} R_\rho} - \frac{\mathrm{d}R_\rho}{R_\rho} \right) \right].
    \label{SIeq: steady-state}
\end{equation}
If we choose
\begin{equation}
    \mathrm{d}R_\rho = \begin{cases}
        \epsilon R_\rho, & \text{if } \rho = \rho^*; \\
        0, & \text{else}
    \end{cases}
\end{equation}
for a specific transition type $\rho^*$ with a small value $\epsilon$, \cref{SIeq: steady-state} becomes \cref{eq: rate function linear response} in the main text.

\bibliography{apssamp}

\end{document}